\documentclass[twocolumn,showpacs,preprintnumbers,amsmath,amssymb,aps,prl]{revtex4-1}

\usepackage{amssymb,amsfonts,amsmath}
\usepackage{euscript}
\usepackage{graphics}
\usepackage{setspace}
\usepackage{color}
\usepackage{graphicx}
\usepackage{hyperref}

\begin{document}

\title{Stability and dynamics of anisotropically-tumbling chemotactic swimmers}

\author{Enkeleida Lushi}
\affiliation{$^1$School of Engineering, Brown University, Providence RI 02912, USA \\
                $^2$ Courant Institute of Mathematical Sciences, New York, NY, 10012, USA}
\email{enkeleida\_lushi@brown.edu}


\begin{abstract}
Micro-swimmers such as bacteria perform random walks known as run-and-tumbles to move up chemo-attractant gradients and as a result aggregate with others. It is also known that such micro-swimmers can self-organize into macroscopic patterns due to interactions with neighboring cells through the fluidic environment they live in. While the pattern formation resulting from chemotactic and hydrodynamic interactions separately and together have been previously investigated, the effect of the anisotropy in the tumbles of micro-swimmers has been unexplored. Here we show through linear analysis and full nonlinear simulations that the slight anisotropy in the individual swimmer tumbles can alter the collective pattern formation in non-trivial ways. We show that tumbling anisotropy diminishes the magnitude of the chemotactic aggregates but may result in more such aggregation peaks.
\end{abstract}



\maketitle

\section{Introduction}
\label{Introduction}

Micro-swimmers such as bacteria \textit{Escherichia coli} perform a biased random walk that enables them to move up regions of increasing chemical they are attracted to \cite{Berg93}. This chemo-attractant is typically food they consume, but can also be chemicals that bacteria signal each-other with \cite{Bassler02} in quorum sensing and communication \cite{ParkEtAl03}. The random walk such swimming bacteria perform, consisting of a sequence of straight runs and tumbles, is biased as the mean run duration increases when a bacterium moves in the direction of the chemo-attractant gradient. As a result, the bacteria eventually aggregate in regions of high chemo-attractant levels. Though for simplicity most mathematical models studying chemotaxis and bacterial random walks assume independence of the pre- and post-tumble directions, in bacteria like {\it E. coli} these directions are in fact correlated \cite{Schnitzer93,SubKoch09}. This slight correlation in the directions, or tumbling anisotropy, can result in a different individual and collective dynamics that has not been previously explored.

Many motile bacteria swim and live in fluidic environments. Their mechanical interactions through this fluid medium can affect their collective self-organization even in the absence of externally-imposed flows, chemical cues or other possible stimuli \cite{DombEtAl04, CisnEtAl07}. If the bacteria are chemotactic, the chemicals they produce or consume can be transported or diffused in the fluid, and hence the modes of communication as well as pattern formation can be affected. In particular, micro-swimmers like bacteria, which propel using rear-mounted flagella and are classified as {\it pushers}, are known to self-organize in structures larger in scale and speed than an individual due to direct collisions and hydrodynamical interactions \cite{DombEtAl04,CisnEtAl07,Sokolov07,SaintShelley08, SubKoch11, CisnEtAl11,Dunkel13, Lushi14}. In a recent study \cite{LGS12}, we found that fully coupling the fluid motion to the dynamics of the swimmers and chemo-attractant can greatly affect and modify the colony's pattern formation.

Here we explore the combined effects of the anisotropic chemotaxis and the collectively-generated fluid flows through linear analysis and nonlinear simulations. The Run-and-Tumble chemotaxis model we use here is based on Alt's work \cite{Alt80}, and subsequent analysis of Schnitzer \cite{Schnitzer93}, Bearon and Pedley \cite{BearonPedley00}, and Chen \textit{et al.} \cite{ChenEtAl03} on a continuum formulation of the biased random walk in three dimensions. In particular, we extend our recent model of chemotactic dynamics in the presence of self-generated fluid flows \cite{LGS12, LGS13} to include an anisotropic run-and-tumble chemotactic response in the theory of motile suspensions. 

We discover that while the   major determinants of the pattern formation are chemotaxis and the collectively-generated fluid flows, the tumble anisotropy still results in subtle but non-trivial alterations in the dynamics. Linear analysis predicts that the tumbling anisotropy can stabilize the chemotaxis-induced concentration growth, though it has little effect on the instability due to hydrodynamic interactions between the swimmers. Full nonlinear simulations of the coupled equations reveal that not only does the tumble anisotropy weaken the chemotactic aggregation, it generally results in more swimmer concentration peaks that are lower in magnitude.

\section{Mathematical Model}
\label{Models}

\subsection{Run-and-tumble Auto-chemotaxis in 3D}

We consider ellipsoidal micro-swimmers each propelling with a constant speed $U_0:=1$ 
in a 3D fluid domain. The swimmer's
center of mass is denoted by $\mathbf{x}$ and its swimming direction $\mathbf{p}$ 
 (with $|\mathbf{p}|=1$) is along the ellipsoid's major axis. We represent the configuration of micro-swimmers by a distribution function
$\Psi(\mathbf{x},\mathbf{p},t)$. The dynamics of a suspensions of swimmers that individually perform run-and-tumble biased walks, is then described by a conservation equation 
\begin{align}
 \frac{\partial \Psi}{\partial t} &= - \nabla_x \cdot [ \Psi \dot{\mathbf{x}} ] - \nabla_p \cdot [ \Psi   \dot{ \mathbf{p}} ] \nonumber \\
 &- [\Psi \lambda (\mathcal{D}_t C)  - \int \mathbf{K}(\mathbf{p},\mathbf{p'};\delta) \Psi  (\mathbf{p'}) \lambda (\mathcal{D}_t C)  d \mathbf{p'}] \label{runandtumble3D} \\
  \dot{\mathbf{x}} &= U_0 \mathbf{p} + \mathbf{u}  - D \nabla_x (\ln \Psi)  \label{xdot} \\
  \dot{ \mathbf{p}} &=  (\mathbf{I}-\mathbf{pp})(\gamma \mathbf{E} + \mathbf{W}) \mathbf{p} - D_r \nabla_p (\ln \Psi). \label{pdot}
\end{align}

Eqs. (\ref{xdot}) and (\ref{pdot}) describe changes
of the swimmer position and orientation. Eq. (\ref{xdot}) says that a swimmer propels itself along its major axis
$\mathbf{p}$ with speed $U_0$ while also being advected by the
 fluid flow $\mathbf{u}$. The last term describes
isotropic translational diffusion with constant $D$.
Eq. (\ref{pdot}) describes the rotation of an ellipsoid by the local background flow (Jeffery's equation),
with $\mathbf{E} = (\nabla \mathbf{u} + \nabla^T \mathbf{u} )/2$,
$\mathbf{W} = (\nabla \mathbf{u} -\nabla^T \mathbf{u} )/2$ the
the rate-of-strain and vorticity tensors, respectively, and $\gamma$ a
shape parameter $-1 \leq \gamma \leq 1$ (for a sphere
$\gamma=0$ and for a rod-like swimmer $\gamma \approx 1$). 
The last term in Eq. (\ref{pdot}) describes the swimmer rotational diffusion with
angular diffusion constant $D_r$, as modeled in recent studies of non-chemotactic swimmers \cite{SaintShelley08b, HohenShelley10, LGS13}.

The run-and-tumble chemotaxis process is described by the second line in Eq. (\ref{runandtumble3D}). 
Runs are assumed mostly straight and the tumbles are assumed mostly instantaneous \cite{Alt80, Schnitzer93, ChenEtAl03}.
Here $\lambda(\mathcal{D}_t C)$ is the tumbling frequency or stopping rate and it is related to the probability of a bacterium having a tumbling event over a fixed time interval. From observations in experiments \cite{MacnabKoshland72}, when the time rate of change of the chemo-attractant gradient is positive along the path of a swimmer, the swimmer's tumbling frequency reduces. If the chemo-attractant concentration is constant or decreasing, the stopping rate is constant. Based on experimental data \cite{MacnabKoshland72} and previous theoretical studies \cite{ChenEtAl03}, this biphasic response can be modeled in a piece-wise continuous way
\begin{eqnarray}\label{stoppingrate}
 \lambda(\mathcal{D}_t C) = \left\{
  \begin{array}{l l}
     \lambda_0 \exp \left( -\chi \mathcal{D}_t C  \right) & \quad \text{if } \mathcal{D}_t C  >0\\
    \lambda_0 & \quad \text{otherwise},
  \end{array} \right.
\end{eqnarray}
where
\begin{align} \label{DcDt} \mathcal{D}_t C 
= \frac{\partial C}{\partial t} 
+ \left( \mathbf{u} + U_0 \mathbf{p} \right) \cdot \nabla{C}
\end{align}
is the rate-of-change of the chemo-attractant concentration along the bacterium's path. The parameter $\lambda_0$ is the basal tumbling frequency (or basal stopping rate) in the absence of chemotaxis, and $\chi$ the chemotactic strength or sensitivity. In literature the response has been approximated in various forms, e.g. exponential as above \cite{ChenEtAl03}, linearized \cite{BearonPedley00}, but typically does not include the temporal chemo-attractant gradient, the swimmer propulsion or its advection by the moving fluid \cite{TindalEtAl08}. Our recent studies \cite{LGS12, LGS13} include both the chemo-attractant and fluid dynamics. 
 
We include the tumbling frequency $\lambda(\mathcal{D}_t C)$ in a linearized piece-wise continuous form of Eq. (\ref{stoppingrate})
\begin{eqnarray}\label{stoppingrate_lin}
 \lambda(\mathcal{D}_t C) = \left\{
  \begin{array}{l l}
     \lambda_0 \left( 1 -\chi \mathcal{D}_t C \right)& \quad \text{if } 0< \mathcal{D}_t C <1/\chi \\
     0 & \quad \text{if } 1/\chi< \mathcal{D}_t C \\
    \lambda_0 & \quad \text{otherwise} 
  \end{array} \right.
\end{eqnarray}

The integral term in Eq. (\ref{runandtumble3D}) includes a ``turning kernel'' $\mathbf{K}(\mathbf{p},\mathbf{p'};\delta)$ which represents a conditional probability of a bacterium tumbling from direction $\mathbf{p}$ to post-tumble direction $\mathbf{p'}$. The parameter $\delta \geq0$ represents a correlation of the pre- and post-tumble directions, motivated by the fact that the tumbles are not perfectly random even in the absence of chemotaxis. Mathematically the anisotropy is dependent on the absolute difference $|\mathbf{p} - \mathbf{p'}|$. A natural choice for the turning kernel dependent on this different is an exponential form $exp \left( \delta (\mathbf{p} \cdot \mathbf{p'}) \right)$. Assuming a tumble happens ($ \mathbf{p'} \neq \mathbf{p}$), the turning kernel has to satisfy a few conditions \cite{Schnitzer93, SubKoch09}:

\begin{itemize}
\item The integral over all directions should equal $1$, thus $\int d\mathbf{p'}\mathbf{K}(\mathbf{p},\mathbf{p'};\delta) = 1$, as the total number of swimmers is conserved. 
\item $\mathbf{K}(\mathbf{p},\mathbf{p'};\delta) \rightarrow 1/4\pi$ as $\delta \rightarrow 0$, so the tumbles should be perfectly isotropic in the case of no correlations between pre- and post-tumble directions.
\item $\mathbf{K}(\mathbf{p},\mathbf{p'};\delta) \rightarrow \Delta(\mathbf{p}-\mathbf{p'})$ (a Dirac delta in orientation) as $\delta \rightarrow \infty$, so each tumble leads to infinitesimally small changes.
\end{itemize}
Such a kernel is proposed by Subramanian and Koch \cite{SubKoch09}
\begin{align}\label{turningkernel3D}
  \mathbf{K}(\mathbf{p},\mathbf{p'};\delta) = \frac{\delta }{4 \pi \sinh(\delta)} e^{\delta \mathbf{p} \cdot \mathbf{p'}} .
\end{align}

For $\delta \rightarrow 0$ we get perfectly isotropic tumbles while in the limit $\delta \rightarrow \infty$ each tumble leads to only infinitesimally small changes in direction, so we get "smoothly-turning swimmers". For \textit{E. coli}, $\delta \approx 1$, as also explained by Subramanian and Koch \cite{SubKoch09}, since their mean angle of tumbling is about $68^o$. Some previous studies, e.g. \cite{BearonPedley00, LGS12, LGS13}, have considered only isotropic tumbles, that is $\delta=0$ and $\mathbf{K}(\mathbf{p},\mathbf{p'};0) =1/4\pi $. We will examine here the effects of this tumbling anisotropy in the dynamics of an active chemotactic suspension.

The fluid velocity $\mathbf{u}(\mathbf{x},t)$ satisfies the non-dimensionalized Stokes equations with an extra or active stress due to the  swimmers' locomotion in it
\begin{align}\label{Stokes-dim}
 - \nabla^2 \mathbf{u} +\nabla q &= \nabla \cdot \Sigma^a \nonumber \\
 \nabla \cdot \mathbf{u} &= 0.
\end{align} 
Here $q$ the fluid pressure and $\Sigma^a$ the active stress   
\begin{align}\label{stress-nondim}
 \Sigma^a (\mathbf{x},t) = \alpha \int \Psi (\mathbf{x},\mathbf{p},t) (\mathbf{pp}^T-\mathbf{I}/3)d\mathbf{p}.
\end{align}
 The active stress is a configuration average over all orientations $\mathbf{p}$ of the stresslets (or force-dipoles) $\alpha (\mathbf{pp}^T-\mathbf{I}/3)$ exerted by the swimmers on the fluid \cite{SaintShelley08}. The stresslet strength $\alpha$ is a $O(1)$ dimensionless constant that depends on the mechanism of swimming and swimmer geometry \cite{SaintShelley08b}. For \textit{pusher} swimmers that propel themselves by exerting a force near the tail, e.g. bacteria like \textit{B. subtilis} or \textit{E. coli}, $\alpha<0$.  For \textit{puller} swimmers that propel using front-mounted flagella, e.g. algae \textit{C. reinhardtii}, $\alpha>0$.

We define the local swimmer concentration $\Phi (\mathbf{x},t)$ as
\begin{align}\label{PhiEqn}
 \Phi(\mathbf{x},t) &= \int \Psi (\mathbf{x},\mathbf{p},t) d \mathbf{p}.
\end{align}

The chemo-attractant is also dispersed in the fluid and has a dynamics of its own that includes advection by the fluid and molecular diffusion. 
 We model the chemo-attractant dynamics as in the original Keller-Segel paper \cite{KellerSegel70,KellerSegel71,LushiDissert} but include fluid advection 
\begin{align}\label{chemo}
 \frac{\partial C}{\partial t} = -\mathbf{u} \cdot \nabla C -\beta_1 C +\beta_2 \Phi + D_c \nabla^2 C.
\end{align}
Here $-\beta_1 C$ is a chemo-attractant degradation term with constant rate $\beta_1$. The term $\beta_2 \Phi$ describes production (for $\beta_2>0$) or consumption ($\beta_2<0$) of the chemo-attractant by the the micro-swimmers. The last term describes spatial diffusion with diffusion coefficient $D_c$. There are two possibilities: the attractant is externally supplied, or alternatively the micro-swimmers themselves produce it. For simplicity of analysis, we focus here on the case of \textit{auto-chemotaxis} $\beta_2>0$ where the swimmers themselves produce the attractant. For studies involving the case of external chemo-attractants, see \cite{Sokolov09,Kasyap12,Ezhilan12,Kasyap14}.

The chemo-attractant equation Eq.(\ref{chemo}), together with the equation for the probability distribution function $\Psi$ in Eq.(\ref{runandtumble3D}) (and hence $\Phi$) and the Stokes Equations with active particle stress Eq.(\ref{Stokes-dim}), describes the dynamics of a swimmer suspension influenced by an anisotropic run-and-tumble chemotaxis in an evolving chemical field. 

For the anisotropic run-and-tumble auto-chemotaxis model in two dimensions, see the Appendix.

\section{Linear Stability Analysis}
\label{Stability}

\subsection{The Eigenvalue Problem}
\vspace{-.1in}
We analyze the linear stability of auto-chemotactic suspensions, $\beta_1,\beta_2>0$ in Eq. (\ref{chemo}), about the uniform and isotropic state $\Psi_0=1/4\pi$. For simplicity, we consider no swimmer diffusion ($D=D_r=0$) and only a quasi-static chemo-attractant field 
\begin{align}\label{quasistaticchem}
 - \beta_1 C + \beta_2 \Phi+ D_c \nabla^2 C =0.
\end{align}

We consider perturbations of the swimmer distribution and chemo-attractant about the uniform isotropic state ($\Psi_0=1/4\pi$) and steady-state ($\overline{C} = \beta_2/\beta_1$) 
\begin{eqnarray}
\Psi(\mathbf{x}, \mathbf{p},t) &= \frac{1}{4\pi} + \epsilon \Psi'(\mathbf{x}, \mathbf{p},t), \quad C(\mathbf{x}, t) = \frac{\beta_2}{ \beta_1} + \epsilon C'(\mathbf{x}, t) \nonumber 
\end{eqnarray}
with $|\epsilon|<<1$. This choice simplifies the stopping rate to
\begin{align}\label{linstoprate}
\lambda(\mathcal{D}_t C) = \lambda_0\left(1- \chi \mathbf{p} \cdot \nabla C  \right).
\end{align}

The linearized equation for the distribution then is 
\begin{align}\label{linPsi1}
\frac{\partial \Psi'}{ \partial t} &= - \mathbf{p}^T \nabla \Psi' + \frac{3 \gamma}{4\pi} \mathbf{p}^T \mathbf{E'}\mathbf{p}  \\
 &-\lambda_0 \Psi'  + \frac{\lambda_0 \chi}{4\pi} \mathbf{p}^T\nabla C' \nonumber \\
 &+ \lambda_0 \int \mathbf{K}(\mathbf{p},\mathbf{p'};\delta) \Psi'( \mathbf{p}') d\mathbf{p}' \nonumber \\
 &- \frac{\lambda_0 \chi}{4\pi} \int \mathbf{K}(\mathbf{p},\mathbf{p'};\delta) \mathbf{p'}^T d\mathbf{p}' \nabla C'. \nonumber
\end{align}

When anisotropic tumbles are included in the chemotaxis model, the turning kernel $\mathbf{K}(\mathbf{p},\mathbf{p'};\delta)$ couples all the rotational moments and makes it difficult to analyze the problem. To get some insight on the role of anisotropy, we look at the linearized turning kernel for  $0 < \delta <1$
\begin{align*}
K(\mathbf{p}, \mathbf{p'}; \delta) \approx \frac{1}{4\pi} + \frac{1}{4\pi} \delta \mathbf{p} \cdot \mathbf{p'}.
\end{align*}

Substituting these into the linearized Eq. (\ref{linPsi1}), and simplifying, we obtain

\begin{align}\label{linPsi2}
\frac{\partial \Psi'}{ \partial t} &= - \mathbf{p}^T \nabla \Psi' + \frac{3 \gamma}{4\pi} \mathbf{p}^T \mathbf{E'}\mathbf{p}  -\lambda_0 \Psi' \\
&+ \frac{ \lambda_0 \chi}{4 \pi} ( 1-\frac{\delta}{3} )  \mathbf{p}^T \nabla C'  + \frac{\lambda_0}{4 \pi} \int \Psi'(\mathbf{p}') d \mathbf{p'} \nonumber \\
&+  \frac{\lambda_0 \delta }{4 \pi} \mathbf{p}^T\int \mathbf{p'} \Psi'(\mathbf{p}')  d \mathbf{p'}   \nonumber
\end{align}

Next, we consider a plane wave perturbation for the distribution function $\Psi'(\mathbf{x}, \mathbf{p},t) = \tilde{\Psi}(\mathbf{k}, \mathbf{p}) \exp (i \mathbf{k}^T\mathbf{x}+\sigma t) $ and other quantities. Here $\mathbf{k}=k\hat{\mathbf{k}}$ is the wavenumber. 

For the quasi-static chemo-attractant form in Eq.(\ref{quasistaticchem}), the chemo-attractant concentration then can be solved in terms of the swimmer concentration $\Phi$, and hence $\Psi$
\begin{align}\label{tildeC}
\tilde{C} = \frac{\beta_2}{\beta_1 + k^2 D_c}\tilde{\Phi} = \frac{\beta_2}{\beta_1 + k^2 D_c}\int \tilde{\Psi}(\mathbf{p'})  d\mathbf{p'} 
\end{align}
where $k= |\mathbf{k}|$. We can solve the fluid equations for the fluid velocity perturbation in terms of the active stress
\begin{align}
\tilde{\mathbf{u}} = \frac{i}{k}(\mathbf{I} - \hat{\mathbf{k}}\hat{\mathbf{k}}^T)  \tilde{\Sigma^a} \hat{\mathbf{k}}.
\end{align}
Since  the active stress is related to the swimmer distribution as
$\tilde{\Sigma^a} = \alpha \int \mathbf{p'p'}^T \tilde{\Psi}( \mathbf{p'}) d\mathbf{p'}$, then
\begin{align}\label{tildeE}
\nabla \tilde{\mathbf{u}} = -\alpha (\mathbf{I} - \hat{\mathbf{k}}\hat{\mathbf{k}}^T) \int \mathbf{p'p'}^T \tilde{\Psi}( \mathbf{p'}) d\mathbf{p'}  \hat{\mathbf{k}}\hat{\mathbf{k}}^T.
\end{align}

Substituting Eqs. (\ref{tildeC},\ref{tildeE}) into Eq. (\ref{linPsi2}), we get a closed equation
for the distribution mode $\tilde{\Psi}$:

\begin{align}
\sigma \tilde{\Psi} &= -i k \mathbf{p}^T \hat{\mathbf{k}} \tilde{\Psi} \nonumber \\
&-\frac{3 \alpha \gamma}{4 \pi} \mathbf{p}^T(\mathbf{I} - \hat{\mathbf{k}}\hat{\mathbf{k}}^T) \int \mathbf{p'p'}^T \tilde{\Psi}( \mathbf{p'}) d\mathbf{p'}  \hat{\mathbf{k}}\hat{\mathbf{k}}^T\mathbf{p} \nonumber \\
&-\lambda_0 \tilde{\Psi} + \frac{\lambda_0 \chi}{4 \pi}(1-\frac{\delta}{3}) \frac{\beta_2 i k}{\beta_1 +k^2D_c} \mathbf{p}^T \hat{\mathbf{k}}  \int \tilde{\Psi}(\mathbf{p'})  d\mathbf{p'} \nonumber \\
&+ \frac{\lambda_0}{4\pi} \int \tilde{\Psi}(\mathbf{p'})  d\mathbf{p'} + \frac{\lambda_0 \delta}{4\pi} \mathbf{p}^T \int \mathbf{p'} \tilde{\Psi}(\mathbf{p'})  d\mathbf{p'}.
\end{align}

Without loss of generality we  let $\hat{\mathbf{k}} = \hat{\mathbf{z}}=[0;0;1]$. Recall that $\mathbf{p}=[\sin \theta \cos \phi; \sin \theta \sin \phi ; \cos \theta]$ and $d \mathbf{p} = \sin \theta d \theta d \phi$ for $\phi \in [0,2\pi)$, $\theta \in [0,\pi]$. Then we can write

\begin{align}\label{Psitilde_delta}
(\sigma &+ \lambda_0 + ik \cos \theta) \tilde{\Psi} =
 \frac{-3 \alpha \gamma}{4 \pi} \cos \theta \sin \theta [ \cos \phi F_1+ \sin \phi F_2] \nonumber \\
&+ \frac{\lambda_0}{4 \pi} \left[ \frac{\chi \beta_2}{(\beta_1 + k^2 D_c)} ik \cos \theta(1- \frac{\delta}{3}) +1\right] G \nonumber \\
&+\frac{\lambda_0 \delta}{4 \pi}   [\sin \theta \cos \phi H_1+ \sin \theta \sin \phi H_2+ \cos \theta H_3] 
\end{align}

where for simplicity we have defined the following integral operators of $\tilde{\Psi}$

\begin{align}
F_1(\tilde{\Psi}) &=  \int_0^{2\pi}  \cos \phi' \int_0^{\pi} \sin^2 \theta' \cos \theta' \tilde{\Psi} (\theta',\phi') d\theta' d \phi' \nonumber \\
F_2(\tilde{\Psi})  &= \int_0^{2\pi}  \sin \phi' \int_0^{\pi}  \sin^2 \theta' \cos \theta' \tilde{\Psi}(\theta',\phi') d\theta' d \phi'\nonumber \\
G(\tilde{\Psi})  &= \int_0^{2\pi}  \int_0^{\pi} \sin \theta' \tilde{\Psi}(\theta',\phi') d\theta' d \phi'\nonumber \\
H_1(\tilde{\Psi})  &=  \int_0^{2\pi}  \cos \phi' \int_0^{\pi}  \sin^2 \theta' \tilde{\Psi}(\theta',\phi') d\theta' d \phi'\nonumber \\
H_2(\tilde{\Psi})  &= \int_0^{2\pi}  \sin \phi' \int_0^{\pi}  \sin^2 \theta' \tilde{\Psi}(\theta',\phi') d\theta' d \phi'\nonumber \\
H_3 (\tilde{\Psi})  &=  \int_0^{2\pi} \int_0^{\pi}  \sin \theta' \cos \theta' \tilde{\Psi}(\theta',\phi')d\theta' d \phi'.
\end{align}

Eq. (\ref{Psitilde_delta}) constitutes a linear eigenvalue problem for the perturbation mode $\tilde{\Psi}$ and the growth rate $\sigma$. 

To obtain the eigenvalue relations, we proceed as in \cite{LGS13} and apply each of the above operators $F_1, H_1, G, H_3$ to $\tilde{\Psi}$ in Eq. (\ref{Psitilde_delta}). The expressions obtained then relate $F_1(\tilde{\Psi}), H_1(\tilde{\Psi}), G(\tilde{\Psi}), H_3(\tilde{\Psi})$ as below
\begin{align}\label{eqsFGH}
F_1 &= \frac{-3 \alpha \gamma}{4} J_1 F_1 + \frac{\lambda_0 \delta}{4} J_2 H_1 \\
H_1 &= \frac{-3 \alpha \gamma}{4} J_2 F_1 +\frac{\lambda_0 \delta}{4} J_3 H_1 \nonumber \\
G &= \frac{\lambda_0}{2}Rik (1-\frac{\delta}{3}) J_4 G +\frac{\lambda_0}{2}  J_5 G + \frac{\lambda_0 \delta}{2}  J_4 H_3 \nonumber \\
H_3 &= \frac{\lambda_0}{2}Rik (1-\frac{\delta}{3}) J_6 G +\frac{\lambda_0}{2} J_4 G + \frac{\lambda_0 \delta}{2}  J_6 H_3\nonumber
\end{align}
where the integrals involved are 
\begin{align}
J_1 &= \int_0^{\pi} \frac{ \sin^3 \theta \cos^2 \theta}{\sigma + \lambda_0 + ik \cos \theta} d\theta \nonumber \\
&=  \frac{2a^3}{ik} -\frac{4a}{3ik} + \frac{(a^4-a^2)}{ik}\log \left(  \frac{a-1}{a+1} \right) \nonumber \\
J_2 &= \int_0^{\pi} \frac{ \sin^3 \theta \cos \theta}{\sigma + \lambda_0 + ik \cos \theta} d\theta 
= -J_1/a \nonumber \\
J_3 &= \int_0^{\pi} \frac{ \sin^3 \theta d\theta}{\sigma + \lambda_0 + ik \cos \theta} 
=  \frac{2a}{ik} + \frac{(a^2-1)}{ik}\log \left(  \frac{a-1}{a+1} \right) \nonumber \\
J_4 &= \int_0^{\pi} \frac{ \sin \theta \cos \theta}{\sigma + \lambda_0 + ik \cos \theta} d\theta 
=  \frac{2}{ik} + \frac{a}{ik}\log \left(  \frac{a-1}{a+1} \right) \nonumber \\
J_5 &=  \int_0^{\pi} \frac{ \sin \theta }{\sigma + \lambda_0 + ik \cos \theta} d\theta = - \frac{1}{ik} \log \left(  \frac{a-1}{a+1} \right) \nonumber \\
 J_6 &= \int_0^{\pi} \frac{ \sin \theta \cos^2 \theta}{\sigma + \lambda_0 + ik \cos \theta} d\theta 
 = -aJ_4. \nonumber
\end{align}
For writing simplicity we have defined $a:=(\sigma + \lambda_0)/ik$ and
$R:=\chi \beta_2/(\beta_1+k^2 D_c)$. 
The equations for $F_2, H_2$ are identical to those for $F_1, H_1$, so are omitted.

Note how the expressions in Eqs. (\ref{eqsFGH}) separate into two groups: $F_1, H_1$ and $G, H_3$.  By combining the first two into one equation for $F_1$ and the last two into one equation for $G$, we obtain two separate dispersion relations for $\sigma(k)$ that are compactly written as: 
\begin{align}
0&=\left( 1+ \frac{3 \alpha \gamma}{4} J_1 \right) \left( 1-\frac{\lambda_0 \delta}{4} J_3 \right)+\frac{\lambda_0 \delta}{4} J_2 \frac{3 \alpha \gamma}{4} J_2 \label{hydro-inst-fullrt} \\
0&=\left( 1- \frac{\lambda_0 \delta}{2} J_6\right) \left(  1- \frac{\lambda_0}{2}Rik (1-\frac{\delta}{3}) J_4 -  \frac{\lambda_0}{2}  J_5 \right) \nonumber \\
&- \frac{\lambda_0 \delta}{4}  J_4
\left( \frac{\lambda_0}{2}Rik (1-\frac{\delta}{3}) J_6 + \frac{\lambda_0}{2} J_4    \right). \label{chem-inst-fullrt}
\end{align}

The tumbling anisotropy parameter $\delta$ appears in \textit{both} Eq. (\ref{hydro-inst-fullrt}) and Eq. (\ref{chem-inst-fullrt}), which we will name the hydrodynamics and auto-chemotactic dispersion relations respectively. The hydrodynamic relation, so-called due to the parameters $\alpha$ and $\gamma$ coming from the terms describing the fluid motion, is also affected by the basic stopping rate $\lambda_0$. The auto-chemotactic relation is unaffected by the hydrodynamics and the swimming mechanism, as evidenced by the lack of parameters $\alpha$ of the dipole strength or $\gamma$ the swimmer shape. The auto-chemotactic relation is of course affected by the chemo-attractant dynamics, as evidenced by the presence of the term $R=\chi \beta_1 / (\beta_2 + k^2 D_c)$ which comes from inverting the quasi-static chemo-attractant equation. 

For isotropic suspensions ($\delta=0$) these expressions reduce to the two separate relations found by Lushi {\it et. al.} \cite{LGS12, LGS13} for auto-chemotactic active suspensions. For isotropic, non-chemotactic, non-tumbling, suspensions ($\delta=0$, $\chi=0$, $\lambda_0=0$), Eq. (\ref{hydro-inst-fullrt}) reduces to the one found and studied by others before \cite{SubKoch09,SaintShelley08, SaintShelley08b, HohenShelley10}.

\subsection{Long-wave asymptotic expansions}
\vspace{-.1in}
The dispersion relations of Eqs. (\ref{hydro-inst-fullrt}) and (\ref{chem-inst-fullrt}) cannot be solved exactly for the growth rate $\sigma$. To get insight in the behavior of the system, we look for long-wave (small $k$) asymptotic solutions. Omitting the details of the lengthy calculation, we obtain the following two branches
\begin{align}\label{hydro-asymp}
\sigma_{H1} &\approx -\lambda_0 + \frac{-\alpha \gamma}{5} + \frac{5}{7 \alpha \gamma} \frac{( 9\alpha \gamma + 22\lambda_0\delta  )}{(3\alpha \gamma + 5 \lambda_0\delta )} k^2 + ... \\
\sigma_{H2} &\approx -\lambda_0 (1- \frac{\delta}{3}) - \frac{3}{5 \lambda_0\delta} \frac{( 6\alpha \gamma + 5\lambda_0\delta  )}{(3\alpha \gamma + 5 \lambda_0\delta )} k^2 + ... .
\end{align}

From Eq. (\ref{hydro-asymp}) we can infer that there is a long-wave instability arising from the hydrodynamics in pusher swimmer suspensions ($\alpha=-1$) with elongated shape ($\gamma \neq 0$). 

The auto-chemotactic relation Eq. (\ref{chem-inst-fullrt}) gives only one branch at small $k$ that still satisfies the integral relations:
\begin{align}\label{chem-asymp}
&\sigma_C\approx   \frac{ \frac{\chi \beta_2}{\beta_1}\lambda_0(1-\delta/3)-1  }{3\lambda_0(1-\delta/3)} k^2 + ... .
\end{align}

This asymptotic solutions look similar in form to the ones for isotropic tumbles discussed in \cite{LGS12, LGS13}. The chemotactic instability in Eq. (\ref{chem-asymp}) tells us the anisotropy has a significant impact for a system of finite size. Anisotropic tumbles overall have a stabilizing effect on the suspension, since the growth rate now is smaller. Specifically, if all other parameters are kept constant, this tells us that the chemotactic sensitivity $\chi$ has to be greater to overcome the tumbling anisotropy. 

From Eq. (\ref{chem-asymp}), we can obtain a range of parameters for which to obtain $\sigma_C>0$ and have a chemotactic instability. That happens for $\chi \beta_2 / \beta_1 > 1 / [ \lambda_0(1-\delta /3)]$. The chemo-attractant diffusion comes in at the next order term in Eq. (\ref{chem-asymp}) and it has a stabilizing effect.

The stability analysis of the two-dimensional system is discussed in the Appendix.

\vspace{-.2in}
\subsection{Solving the Dispersion Relation}
\vspace{-.15in}
We solve numerically the dispersion relations in Eqs. (\ref{hydro-inst-fullrt}, \ref{chem-inst-fullrt}) for $\sigma(k)$ by using an iterative solver. To ensure that we do not get spurious solutions, we make use of the asymptotic expansions in Eqs. (\ref{hydro-asymp}) and (\ref{chem-asymp}) as initial guesses for small $k$. Then we solve for $\sigma_k$ for each increasing $k$ and use the previous $k$ solution as an initial guess. The numerical solutions are checked that they still satisfy the integral relations versions of Eqs. (\ref{hydro-inst-fullrt},\ref{chem-inst-fullrt}).

\begin{figure}[htps]
\vspace{-.in}
\centering
\includegraphics[width=0.7\columnwidth]{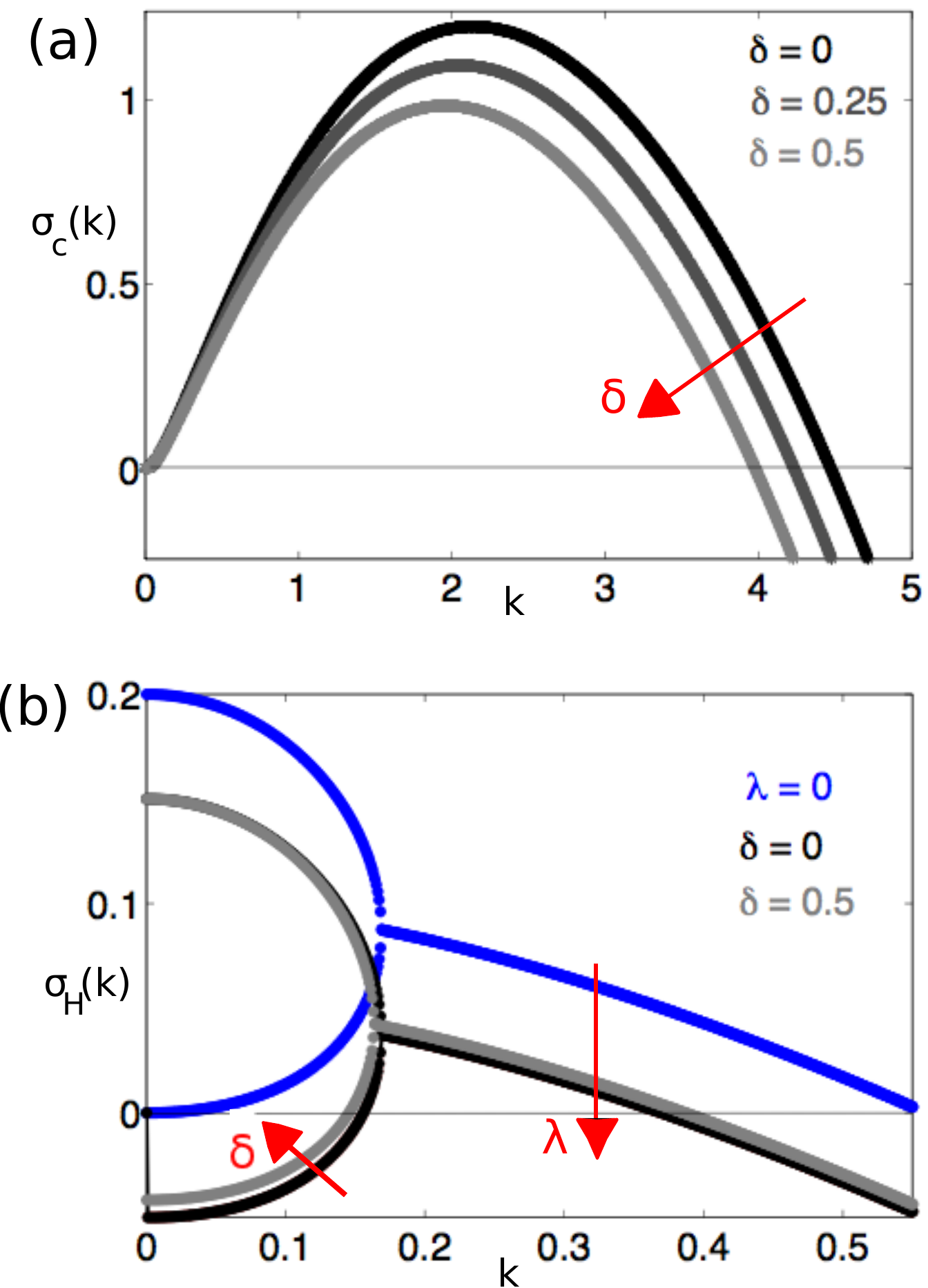}
\vspace{-.1in}
\caption{ (color online) (a) Numerical solution for $\sigma_C(k)$ of the auto-chemotactic relation Eq. (\ref{chem-inst-fullrt}) for $\chi=20$, $D_c=1/20$, $\lambda=0.25$ and a variety of tumbling anisotropy parameters $\delta=0, 0.25, 0.5$. (b) Numerical solution  for $\sigma_H(k)$ of the hydrodynamic relation Eq. (\ref{hydro-inst-fullrt}) for pushers $\alpha=-1$ for $\lambda_0=0, 0.05$ and tumbling anisotropy parameters $\delta=0, 0.5$. Red arrows show the effect of an increasing parameter.
\vspace{-.1in}
}\label{fig:GrowthRates}
\end{figure}

Long-wave asymptotics on the auto-chemotaxis dispersion relation given in Eq. (\ref{chem-inst-fullrt}) gives that for $(\chi \beta_2/\beta_1) \lambda_0 > (1-\delta/3)$ there are wavenumbers with $Re(\sigma_C(k))>0$, for pushers and pullers alike and any shape parameter $\gamma$. Auto-chemotaxis introduces an instability branch, which is solved numerically from Eq.(\ref{chem-inst-fullrt}) and plotted in Fig.\ref{fig:GrowthRates}a. From the plots we see that the tumbling anisotropy $\delta$ can have a visible effect on the growth rates, as also expected from the small $k$ analysis. The range of wavenumbers with $Re(\sigma_C(k))>0$ is smaller for $\delta>0$.

The solution to the hydrodynamics relation is shown in Fig.\ref{fig:GrowthRates}b for rod-like $\gamma=1$ for tumbling pusher swimmers $\alpha=-1$ with basic stopping rate $\lambda_0=0.05$ and cases $\delta=0, 0.5$. The branch $\lambda_0=0.05$ and $\delta=0$ is exactly that obtained by Lushi {\it et al} \cite{LGS12, LGS13} for  swimmers with uncorrelated tumbles. The addition of a small tumbling anisotropy $\delta=0.5$ has barely a visible effect on $\sigma_H$. 

In the case of pullers $\alpha=-1$, there is no hydrodynamic instability as $Re(\sigma_H(k))<0$ for any $\lambda_0$ and $\delta$.

For non-tumbling pushers ($\alpha<0$) there is a hydrodynamic instability for a finite band of wavenumbers $k=0$ until $k_c \approx 0.55$ \cite{SaintShelley08b,HohenShelley10}. Tumbling diminishes this range of unstable wave-numbers since both branches are brought down by $\lambda_0$. As noted in Refs. \cite{LGS12, LGS13}, $\lambda_0 \geq 0.2$ turns off the hydrodynamic instability for any system size, for any swimmer shape $\gamma$, and, as can be surmised from the plot in Fig. \ref{fig:GrowthRates}, any tumbling anisotropy parameter $\delta$.

\vspace{-.15in}
\subsection{Phase Space}
\vspace{-.15in}
Linear theory shows that there is a range of $\lambda_0$ for
which there is a hydrodynamic instability in pusher suspensions. If $\lambda_0
\geq 0.2$, there is no hydrodynamic instability for any system size
and any swimmer shape $\gamma$, since, as seen in
Fig.~\ref{fig:GrowthRates}, $Re(\sigma_H(k))\leq 0.2$. For an
auto-chemotactic instability we need $\chi \beta_2 / \beta_1 > 1 / [ \lambda_0(1-\delta /3)]$. 

\begin{figure}[htps]
\vspace{-.1in}
 \centering
\includegraphics[width=0.7\columnwidth]{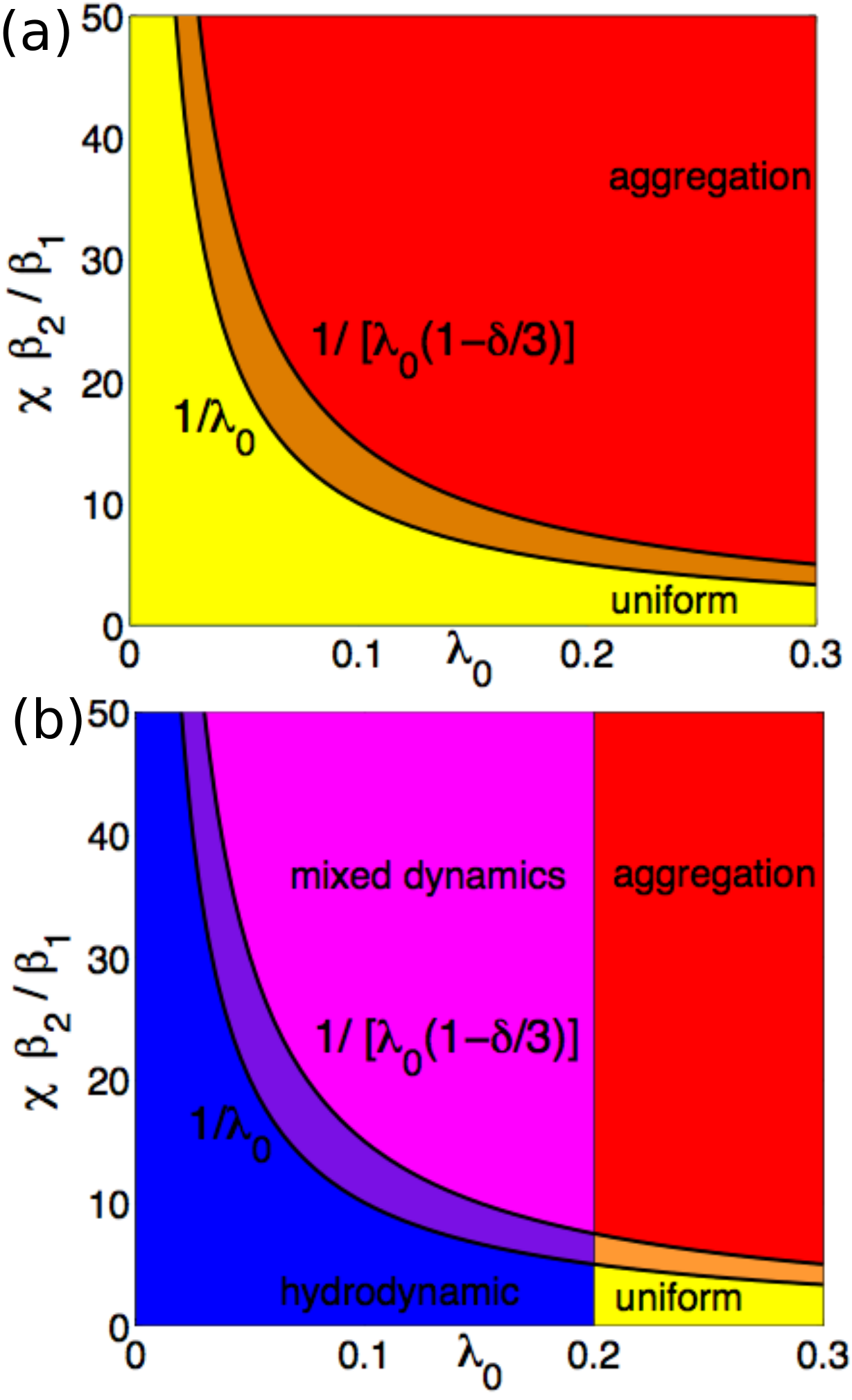}
\vspace{-.1in}
 \caption{ (Color online) Phase space of various regimes for
   auto-chemotactic and/or hydrodynamic instabilities in suspensions of 
   (a) pullers or neutral swimmers, (b) pushers as a function of the basal tumbling frequency $\lambda_0$, correlation of tumbles parameter $\delta$ and chemotactic parameters  $\chi, \beta_1, \beta_2$. The curve $1 / [ \lambda_0(1-\delta /3)]$ is shown for $\delta=0$ and $\delta=1$. 
   }
   \vspace{-.1in}
 \label{fig:phasespace}
 \end{figure}

This connects the auto-chemotaxis parameters $\chi$, $\beta_1$,
$\beta_2$ to the basal tumbling rate $\lambda_0$ and correlation of tumbles parameter $\delta$. This information
about the parameters is assembled in a phase diagram in Fig.~\ref{fig:phasespace}, which
shows the dynamical regimes we expect
based on the linear analysis and nonlinear simulations.

\begin{figure*}[ht]
\centering
\includegraphics[width=1.75\columnwidth]{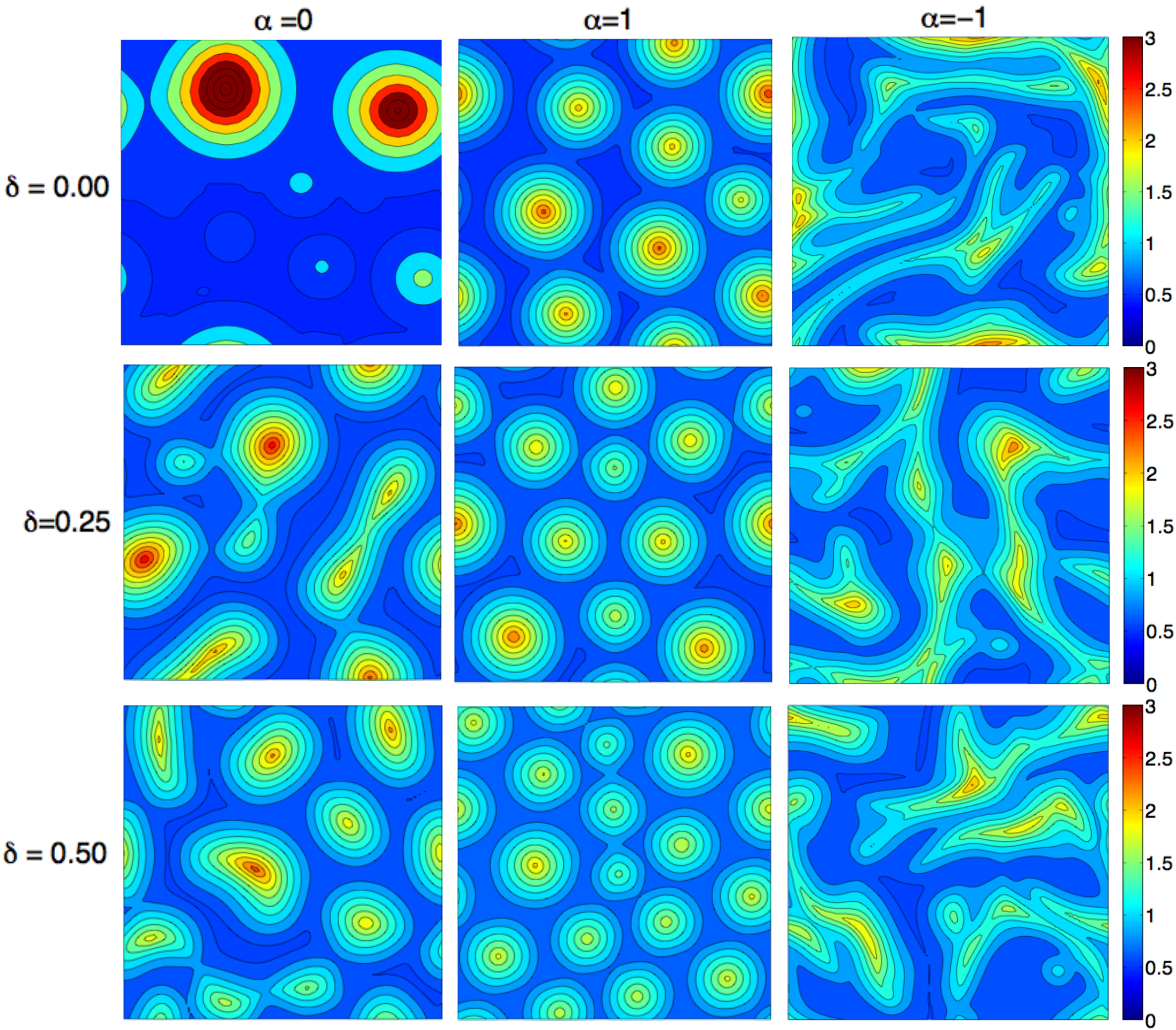}
\vspace{-0.in}
\caption{  (Color online)  Snapshot of the dynamics of micro-swimmers concentration field $\Phi$ at long times. Shown are the cases of ``neutral'' swimmers  ($\alpha=0$), pullers ($\alpha=+1$), and pushers ($\alpha=-1$) with isotropic tumbling ($\delta=0$) or with tumbling direction correlation parameters ($\delta=0.25,0.5$). The chemo-attractant dynamics follows closely that of the swimmer concentration.
}
\vspace{-0.in}
\label{fig:dynamics}
\end{figure*}

\section{Nonlinear Simulations}
\label{Simulations}

We numerically simulate the full nonlinear system describing active suspensions under the influence of anisotropic chemotaxis. In 3D the kinetic model involves five configuration variables, three spatial and two angles, making the simulations computationally expensive. As in Refs.\cite{LGS12,LGS13}, for simplicity we constrain the swimmers in the $(x,y)$-plane ($z=0)$ with direction parametrized by only an angle $\theta \in [0, 2\pi)$ so that the direction is $\mathbf{p} = (\cos \theta, \sin \theta, 0)$. The distribution function $\Psi$ is invariant along the $z$-direction: $\Psi(\mathbf{x}, \mathbf{p},t)=\Psi(x,y,\theta,t)$.

All the variables are periodic in $x,y$ and $\theta$ directions, so we use of the fast Fourier transforms to do the all the differentiations and to solve the for the fluid flow in Eq. (\ref{Stokes-dim}).
Integrations in $\theta$ to obtain the swimmer density $\Phi$ in Eq. (\ref{PhiEqn}) and active particle stresses $\Sigma^a$ in Eq.(\ref{stress-nondim}) are done using a trapezoidal rule. $128-256$ points are used in the $(x,y)$ directions and $32-64$ in the $\theta$ direction. 

The conservation equation Eq. (\ref{runandtumble3D}) and the chemo-attractant equation Eq. (\ref{chemo}) are marched in time using a second order scheme. Swimmer translational and rotational diffusions as well as chemo-attractant diffusion are included in all the simulations with typical values of $D=D_r=0.025$ and $D_c=0.05$. All the results we present here are for elongated rod-shaped swimmers with $\gamma=1$ and the spatial square box side is $L=50$. The initial swimmer distribution is taken to be a uniform and isotropic suspension perturbed as
\begin{align}\label{Psi_init}
 \Psi(\mathbf{x},\theta,0) = \frac{1}{2 \pi} \left[ 1+ \sum_i \epsilon_i \cos (\mathbf{k}_i \cdot \mathbf{x} + \xi_i)P_i(\theta) \right],
\end{align}
where $\epsilon_i$ is a random small coefficient ($|\epsilon_i| <0.01$), $\xi_i$ is a random phase and $P_i(\theta)$ is a third order polynomial of $\sin \theta$ and $\cos \theta$ with randomly-chosen $O(1)$ coefficients. The initial chemo-attractant distribution is taken to be uniform $C(\mathbf{x},0)=\beta_2/\beta_1 $.

\subsection{Dynamics: Qualitative Comparisons}\label{dynamics}

We look at the nonlinear dynamics for all swimmer types (neutral with $\alpha=0$, pullers  $\alpha=+1$, pushers $\alpha=-1$), when there is no tumbling anisotropy ($\delta=0$) and two cases of slight tumbling anisotropy ($\delta=0.25$ and $\delta=0.5$). We pick parameters $\chi=20, \beta_1=\beta_2=1/4$ and $\lambda_0=0.5$ that lie in the aggregation regime of all swimmer types in the phase spaces of Fig. \ref{fig:phasespace}. Snapshots of the swimmer concentration are shown in Fig. \ref{fig:dynamics}.

As explored in our previous studies \cite{LGS12, LGS13}, isotropic suspensions ($\delta=0$) of neutral, puller, and pusher swimmers differ in their pattern morphology. For neutral swimmers ($\alpha=0$) where the fluid flows are not taken into account, the observed dynamics is that of continuous aggregation into large peaks. For puller swimmers ($\alpha=+1$), aggregation into stable peaks also occurs, but these peaks are smaller and circular due to the generated straining fluid flows that keep them from merging further \cite{LGS12, LGS13}. For pushers ($\alpha=-1$) we observe dynamic aggregation of the swimmers into irregular peaks that continuously move, merge and break apart. This effect is due to the collectively-generated fluid flows that are known to occur even in the absence of chemotaxis \cite{SaintShelley08b}, but here these flows are able to transport the chemo-attractant field as well as the swimmers themselves and thus can affect the collective chemotactic dynamics \cite{LGS12, LGS13}.

The anisotropy in the tumbling directions has an interesting effect in the swimmers' collective chemotactic dynamics. Linear stability predicts that increasing tumbling correlation parameter $\delta$ will dampen the chemotactic instability, but have no visible impact in the hydrodynamic instability, as seen from asymptotic results [ Eqs. (\ref{hydro-asymp} and \ref{chem-asymp})] and illustrated in Fig. \ref{fig:GrowthRates}. 

The main dynamics is still determined by the type of swimmer (neutral, puller, pusher), however some differences are clearly visible. Most notably, the tumbling anisotropy has affected the number of the resulting aggregation peaks. For example, for isotropic suspensions ($\delta=0$), the neutral swimmer suspension shown in the example of Fig. \ref{fig:dynamics} has $\approx 2$ peaks and the puller suspension has $\approx 9$ peaks. With slight tumbling anisotropy ($\delta=0.25$), the number of peaks in the neutral swimmer suspension has increased to $\approx 6$ and in the puller to $\approx 11$. Higher tumbling anisotropy ($\delta=0.5$) further increases the peak numbers to $\approx 10$ in the neutral swimmer case and $\approx 16$ peaks in the puller swimmer case.

Curiously, the pusher suspension in Fig. \ref{fig:dynamics} does not seem to be visibly affected when the tumbling anisotropy parameter $\delta$ is increased. The collective dynamics of the pusher swimmers is still typified by dynamic aggregation into peaks that continuously merge, move, and then break apart. The height of these peaks is not visibly affected much.

\subsection{Dynamics: Quantitative Comparisons}\label{comparison}
\vspace{-.15in}

\begin{figure}[htps]
 \centering
\includegraphics[width=0.75\columnwidth]{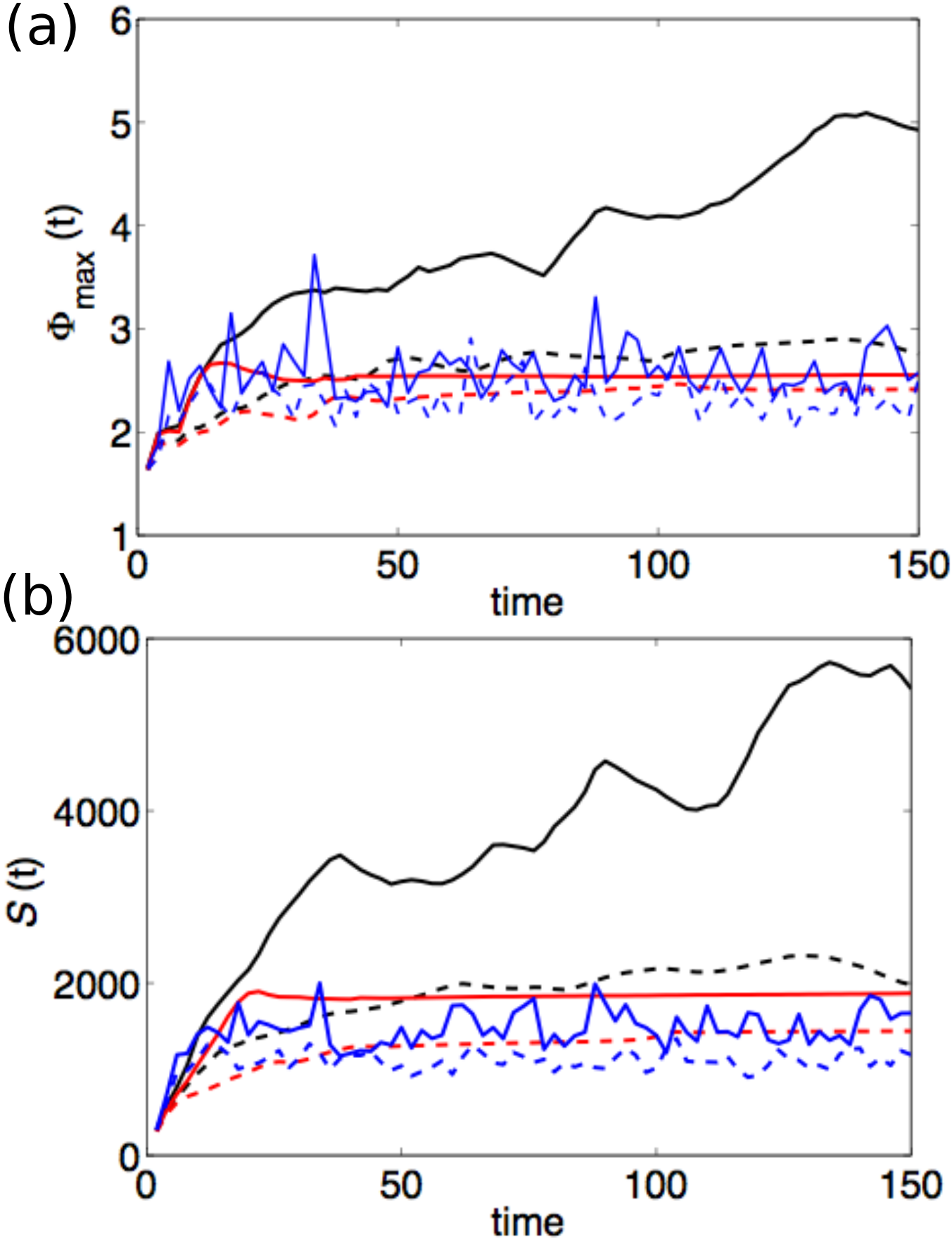}
\vspace{-.1in}
 \caption{ (Color online) Evolution of the maximum of the swimmer concentration $\Phi$ and the system configurational entropy ${\it S}(t)$ in time. The cases shown are for neutral (black line), puller (red line) and pusher swimmers (blue line) for isotropic suspensions (solid lines) and anisotropic suspensions with $\delta=0.25$ (dashed lines). }
   \vspace{-.05in}
 \label{fig:comparison}
 \end{figure}

To quantify the effect of the tumbling anisotropy in the various suspensions, we track the evolution in time of the swimmer concentration maximum $\Phi_{max}$ and the so-called configurational entropy \cite{SaintShelley08, SaintShelley08b, HohenShelley10}
\begin{align}\label{entropy}
 \cal{S}(\mathbf{x}, \mathbf{p}) &= \int \int \Psi (\mathbf{x},\mathbf{p}) \log ( \Psi (\mathbf{x},\mathbf{p}))d \mathbf{p} d \mathbf{x}
\end{align}
which plays the role of a system energy \cite{SaintShelley08b}.

The results, shown in Fig. \ref{fig:comparison}, show that while the tumbling anisotropy does indeed have a dampening effect on the swimmer suspension dynamics which may not be visible in all the snapshots of the dynamics in Fig. \ref{fig:dynamics}. The maximum concentration for anisotropic suspensions, be that of neutral, pusher or puller swimmers, is lower than in the isotropic suspension cases with otherwise the same parameters and initial conditions. Indeed, even in the pusher case, the maximum swimmer concentration is over time lower in the anisotropic suspension. 

The configurational entropy tells us the same story: the tumbling anisotropy dampens the dynamics in all the swimmer suspensions as the lines for the anisotropic cases are lower than the isotropic cases over long times. The major differences in the dynamics still arise due to the swimmer type. 

Qualitatively similar results were obtained for other basal tumbling $\lambda_0$ and chemotactic parameters $\chi$ that were investigated but are not shown here.

\section{Discussion and Conclusion}
\label{Conclusion}

It is well-known that bacteria, after which the so-called pusher swimmers are modeled, perform a run-and-tumble motion even in the absence of chemotaxis \cite{Berg93, BergBrown72}, and in some fashion so do micro-algae like {\it C. reinhardtii}, after which the puller swimmers are modeled \cite{Polin09}. To move up a chemo-attractant gradient, bacteria are known to modify their tumbling rate in response to the local attractant concentration\cite{MacnabKoshland72} and bacteria like {\it E. coli} are known to aggregate in complex and intricate patterns \cite{BudreneBerg, MBBvO03}. These experiments have inspired many theoretical and computational studies of chemotaxis in particular \cite{KellerSegel70, KellerSegel71,Schnitzer93, SchnitzerEtAl90,TindalEtAl08} and micro-swimmer dynamics in general \cite{BearonPedley00,HillPedley05, PedleyKessler92}. However, it is also known that the tumbles in bacteria like {\it E. coli} are not completely random since the pre- and post-tumble directions are slightly correlated \cite{Schnitzer93, SubKoch09,SubKoch11}. The effect of such anisotropy in the run-and-tumble chemotaxis and collective dynamics has barely been explored \cite{SubKoch09,SubKoch11}.

We investigated analytically and computationally the role of correlated tumbles in various micro-swimmer suspensions. We considered the  dynamics resulting from the full coupling of the anisotropic run-and-tumble chemotaxis to the motion of the immersing fluid and the chemo-attractant that the swimmers produce. The types of swimmers considered here are pushers (like swimming bacteria) and pullers (like micro-algae) that are known to individually and collectively disturb the surrounding fluid and affect the neighbors' motion \cite{CisnEtAl07,Sokolov07, DombEtAl04, SaintShelley08, SubKoch09, SubKoch11}, and theoretical neutral swimmers that do not create any fluid disturbances. While neutral and puller swimmers are known to accumulate in peaks due to auto-chemotaxis, pushers dynamically aggregate into aggregates that as a result of the collectively generated fluid flows \cite{LGS12,LGS13}. 

Linear analysis of the system revealed that correlated tumbling affects chemotactic aggregation in all types of swimmers alike and has a stabilizing effect. An instability due to hydrodynamics occurs only in pusher swimmers, and linear analysis predicted that the effect of the correlated tumbles in that case is minor. However, simulations of the full coupled system showed subtle but non-trivial effects of the tumbling anisotropy in the pattern formation. The tumbling anisotropy is predictably a stabilizer on the chemotactic growth of the aggregates in all types of swimmer suspensions. Unpredicted by linear analysis, the tumbling anisotropy is most visibly manifested in the increased number of stable aggregate peaks in suspensions of neutral and puller swimmers. The aggregates observed in the long-time dynamics of the anisotropically-tumbling swimmer suspensions are on average weaker than those in the isotropically-tumbling swimmer suspensions. 

While this study considered the collective motion of anisotropically-tumbling chemotactic swimmers in a fluidic environments, it did not include direct or steric swimmer interactions. Recent simulations and experiments have elucidated the roles of hydrodynamics and shape-specific swimmer collisions in the pattern formation of bacterial suspensions, and neither of these effects are negligible \cite{Lushi14, Wioland16}. Including direct shape-specific collision interactions in continuum theories however is non-trivial, as shown by recent work on rod-shaped non-tumbling non-chemotactic swimmer suspensions \cite{Ezhilan13}. Moreover, it is not yet clear how the swimmer tumbling rate or chemotactic motion are affected by the swimmer density, and how this can be correctly modeled.

We explored analytically and computationally the effect of anisotropic tumbles, as known to occur in the motion of bacteria {\it E. coli}, in the dynamics of various motile suspensions. We hope it leads to an increased interest in studies of such chemotactic micro-swimmers and other active micro-particles \cite{Zoettl16, BenAmar16}.

\section*{Acknowledgments}
\vspace{-0.2in}
The author thanks R. Goldstein, C. Hohenegger and M. Shelley for helpful discussions and gratefully acknowledges funding from NSF Grant NO. CBET-1544196. 
\vspace{-0.2in}


\section{Appendix}
\subsection{The model and linear stability in 2D}

We briefly mention how the 2D system differs from the 3D one. In 2D there is only one orientation angle $\theta \in [0, 2\pi]$ with $\mathbf{p}= (\cos \theta, \sin \theta) $, and the differences from the 3D system are only in the following 
\begin{align}
 \frac{\partial \Psi}{\partial t} &= - \nabla_x \cdot ( \Psi \dot{\mathbf{x}} ) - \partial_{\theta} ( \Psi \dot{\theta} ) \nonumber \\
 &+ \left[\Psi \lambda (\mathbf{p})   
 - \int_0^{2\pi} \mathbf{K}(\mathbf{p},\mathbf{p'};\delta)\Psi' \lambda (\mathbf{p}')  d \theta \right]  \nonumber \\
\dot{\theta} &= \mathbf{p}_{\perp} \cdot (\gamma \mathbf{E} + \mathbf{W}) \mathbf{p}  - D_r \partial_{\theta}(\ln{\Psi} )  \nonumber \\
 \Sigma^a &= \alpha \int_0^{2\pi} \Psi (\mathbf{x},\theta,t) (\mathbf{pp}-\mathbf{I}/2)d\theta
\end{align}
where $\mathbf{p}_{\perp} = (-\sin \theta, \cos \theta)$ is the unit vector perpendicular to the particle orientation. In 2D the isotropic suspension state is given by $\Psi_0=1/2\pi$.

The turning kernel satisfying all conditions in 2D is 
\begin{align}\label{turningkernel_2D}
  \mathbf{K}(\mathbf{p},\mathbf{p'};\delta) = \frac{ 1 }{2 \pi I_0(\delta)} e^{\delta \mathbf{p} \cdot \mathbf{p'}}
\end{align}
with $I_0(\delta)$ a Modified Bessel function of the First Kind.

As with the 3D system, we analyze the linear stability of the 2D system about the uniform isotropic swimmer distribution ($\Psi_0=1/2\pi$) in the case with quasi-static chemo-attractant dynamics and linearized tumbling rate. Two dispersion relations can be obtained here as well, one related to hydrodynamics and the other related to run-and-tumble auto-chemotaxis. 

The long-wave (small $k$) asymptotic analysis of these yields the following growth rates from the hydrodynamic and the auto-chemotactic dispersion relations 
\begin{align}
\sigma_{H1} &\approx -\lambda_0 -\alpha \gamma/4 + O(k^2) \label{hyd1asymp2D} \\
\sigma_{H2} &\approx -\lambda_0 (1-\delta/2) - O(k^2) \label{hyd2asymp2D} \\
\sigma_C &\approx   \frac{ \frac{\chi \beta_2}{\beta_1}\lambda_0(1-\delta/2)-1  }{2\lambda_0(1-\delta/2)} k^2 + O(k^3) \label{chemasymp2D} 
\end{align}
which look qualitatively similar to the 3D dispersion relations. 

As now expected, there are two branches for the growth rate for the hydrodynamic dispersion relation. From Eq. (\ref{hyd1asymp2D}) we can see that there is a hydrodynamic instability only for pusher swimmers ($\alpha=-1$) with elongated shape $\gamma \neq 0$. The tumbling anisotropy does not appear in the highest order terms of the dominant branch $\sigma_{H1}$. Although the tumbling anisotropy appears in the lesser branch $\sigma_{H2}$, this branch stays negative, as also seen in the numerical solution of the 3D analogue in Fig. \ref{fig:GrowthRates}.

Eq. (\ref{chemasymp2D}) shows there is an auto-chemotactic instability for all types of swimmer (neutral, puller and pusher) if the involved parameters satisfy the condition $\chi \beta_2/\beta_1>1/\lambda_0(1-\delta/2)$ which is very similar to its 3D analog. 

.

\end{document}